\documentclass[aps,prd,reprint,preprintnumbers,floats,epsfig,nofootinbib,amssymb,
nofootinbib]{revtex4}

\usepackage{slashed}
\usepackage{graphicx,color}
\usepackage{epsfig}
\usepackage{subfigure}
\usepackage{epsfig}
\usepackage{dcolumn}
\usepackage{bm}
\usepackage{color}
\usepackage{ulem}
\usepackage{enumitem}
\usepackage[colorlinks,
            linkcolor=black,
            anchorcolor=black,
            citecolor=black
            ]{hyperref}

\begin{document}

\baselineskip=15pt


\title{ $\theta_{23}=\pi/4$ and $\delta=-\pi/2$ In Neutrino Mixing, Which Convention? }

\author{Junxing Pan$^{1}$\footnote{panjunxing2007@163.com}}
\author{Jin Sun${}^{2}$\footnote{019072910096@sjtu.edu.cn}}
\author{Xiao-Gang He${}^{3,4,1,2}$\footnote{hexg@phys.ntu.edu.tw}}

\affiliation{${}^{1}$School of Physics and Information Engineering, Shanxi Normal University, Linfen 041004, China}
\affiliation{${}^{2}$Tsung-Dao Lee Institute, and School of Physics and Astronomy, Shanghai Jiao Tong University, Shanghai 200240, China}
\affiliation{${}^{3}$Department of Physics, National Taiwan University, Taipei 10617, Taiwan}
\affiliation{${}^{4}$Physics Division, National Center for Theoretical Sciences, Hsinchu 30013, Taiwan}

\begin{abstract}
Considerable information has been obtained about neutrino mixing matrix. Present data show that in the particle data group (PDG) parameterization, the 2-3 mixing angle and the CP violating phase are consistent with $\theta_{23} = \pi/4$ and $\delta_{PDG} = -\pi/2$, respectively. A lot of efforts have been devoted to constructing models in realizing a mixing matrix with these values. However, the particular angles and phase are parameterization convention dependent. The meaning about the specific values for mixing angle and phase needs to be clarified. Using the well known 9 independent ways of parameterizing the mixing matrix, we show in detail how the mixing angles and phase change with conventions even with the 2-3 mixing angle to be $\pi/4$ and the CP violating phase to be $-\pi/2$. The original Kaobayashi-Maskawa and an additional one belong to such a category. The other 6 parameterizations have mixing angles and phase very different values from those in the PDG parameterization although the physical effects are the same. Therefore one should give the specific parameterization convention when making statements about values for mixing angles and phase.
\end{abstract}

\maketitle

\noindent
{\bf 1. Introduction}

Neutrino oscillation experiments have confirmed that neutrinos have none-zero masses and they mix with each other. With the minimal particle contents in standard model (SM), neutrino masses are zero if the theory is required to be renormalizable. A none-zero neutrino mass, therefore, is an evidence that there are new physics beyond the minimal SM. Precise information about neutrino masses and their mixing pattern can provide crucial information to understand new physics beyond the SM.  The mixing matrix describing neutrino mixing is the Pontecorve-Makai-Nagakawa-Sakata~\cite{PMNS}  mixing matrix $V_{PMNS}$ defined as 
\begin{eqnarray}
L = - {g\over \sqrt{2}} \bar e_i V^{ij}_{PMNS} \gamma^\mu L \nu_j W^{-}_\mu + H.C.\;,
\end{eqnarray}
where $i$ and $j$ are the flavor indices.

For three generations of left-handed neutrinos, in a convention independent manner, the mixing matrix is usually written as
\begin{eqnarray}
V_{PMNS} = \left (
\begin{array}{ccc}
V_{e 1}&\;\;V_{e 2}&\;\;V_{e 3}\\
V_{\mu 1}&\;\;V_{\mu 2}&\;\;V_{\mu 3}\\
V_{\tau 1}&\;\; V_{\tau 2}&\;\; V_{\tau 3}
\end{array}
\right )\;,
\end{eqnarray}
It is a unitary matrix. The specific parameterization in terms of mixing angles and CP violating phase
is convention dependent. There are many different ways to parameterize this mixing matrix.
The popular one used by the Particle Data Group (PDG)~\cite{PDG, maiani, keung} is given by
\begin{eqnarray}
V_{PDG} = \left (
\begin{array}{ccc}
c_{12}c_{13}&\;\;s_{12}c_{13}&\;\;s_{13}e^{-i\delta_{PDG}}\\
-s_{12}c_{23} - c_{12}s_{23}s_{13}e^{i\delta_{PDG}}\;\;& c_{12}c_{23} - s_{12}s_{23}s_{13}e^{i\delta_{PDG}}&\;\;c_{13}s_{23}\\
s_{12}s_{23} -c_{12}c_{23}s_{13}e^{i\delta_{PDG}}&\;\;-c_{12}s_{23}- s_{12}c_{23}s_{13}e^{i\delta_{PDG}}&\;\;c_{13}c_{23}
\end{array}
\right )\;,
\end{eqnarray}
where $s_{ij} = \sin\theta_{ij}$ and $c_{ij} =\cos\theta_{ij}$. There are additional phases depending on if neutrinos are Dirac or Majorana particles which can be lumped into the neutrino masses $\vert m_i\vert exp(i\alpha_i/2)$. For Dirac neutrinos, $\alpha_i = 0$. For Majorana neutrinos, two of the phases (Majorana phases) $\alpha_i$ are physical  which cannot be removed by redefining phases of neutrino fields. We will use the notation that the Majorana phases are in the mass with $m_i =\vert m_i\vert exp(i\alpha_i/2)$, so that the mixing matrix is just parameterized by $\theta_{ij}$ and $\delta_{PDG}$.

Considerable information about the mixing pattern has been obtained experimentally. For the relative size of the three neutrino masses, $m_{i = 1,2,3}$, solar neutrino oscillation data show that $\vert m_1\vert < \vert m_2\vert$. Current data still allow two types of neutrino mass hierarchies, namely, the normal hierarchy (NH) with $\vert m_3\vert  > \vert m_2\vert > \vert m_1\vert$, and the inverted hierarchy (IH) with $\vert m_2\vert > \vert m_1\vert > \vert m_3\vert$. There are some differences for the mixing angles $\theta_{ij}$ and CP violating phase $\delta_{PDG}$ corresponding to these two hierarchies~\cite{PDG}. The best values, and  the $3\sigma$ or 2$\sigma$ ranges for the mixing angles and CP violating phase are given by~\cite{PDG}
\begin{eqnarray}
&&NH: s^2_{12} = 0.297\;, \;\;\;( 3\sigma: 0.250 - 0.354)\;,\;\;\; \;s^2_{23} = 0.425\;,\;\;\;\;\;\;\;\;( 3\sigma: 0.381 - 0.615)\;,\nonumber\\
&&\;\;\;\;\;\;\;\;\;s^2_{13} = 0.0215\;, \;( 3\sigma: 0.0190 - 0.0240)\;,\;\delta_{PDG}/\pi = 1.38\;,\;\; (2\sigma : 1.0 - 1.9)\;.\nonumber\\
&&IH:  \;s^2_{12} = 0.297\;, \;\;\;(3\sigma: 0.250 - 0.354) \;,\;\;\; \;s^2_{23} = 0.589\;, \;\;\;\;\;\;\;\;(3\sigma: 0.384 - 0.636)\;,\\
&&\;\;\;\;\;\;\;\;\;s^2_{13} = 0.0216\;, \;(3\sigma: 0.0190 - 0.0242)\;,\;\delta_{PDG}/\pi = 1.31\;,\;\; (2\sigma: 0.92 - 1.88)\;.\nonumber
\end{eqnarray}
With more data becoming available in the future, the error bars will be improved.

The current data allow the intriguing possibility that, for both NH and IH cases, $\theta_{23}$ and $\delta_{PDG}$ to be $\pi/4$ and $-\pi/2$ (or $3\pi/2$), respectively. Although the central values are not exactly these ones, they can serve as a good approximation. The simple form of the resulting mass matrix and associated symmetry structure have attracted a lot of efforts to understand the underlying reasons and related phenomenology~\cite{GL-symmetry, xing-review, werner,he1,he2,ma1,indian,mohapatra,yasue,ding,zhang}. 
With such values, the neutrino mass matrix $M_\nu$, with appropriate redefinition of phases of charged leptons, takes the following special form for neutrinos being Majorana particles,
\begin{eqnarray}
M_\nu =V_{PDG}\cdot \hat M_\nu\cdot V^T_{PDG}\;, \label{mass}
\end{eqnarray}
where $\hat M_\mu = diag(m_1,m_2,m_3)$ is a diagonal matrix ($diag(a,b,c)$ means a diagonal matrix with the diagonal entries given by a, b and c). In general $m_i$ have Majorana phases. The elements of $M_{\nu}$ are given by~\cite{he1}
\begin{eqnarray}
&&M_{11} = m_1 c_{12}^2 c_{13}^2 +m_2 s_{12}^2 c_{13}^2 - m_3s^2_{13}\;,\nonumber\\
&&M_{22, 33} = {1\over 2} \left (m_1(s^2_{12} - c^2_{12} s_{13}^2) + m_2(c^2_{12} - s^2_{12}s^2_{13}) + m_3 c^2_{13}\right ) \pm i (m_2-m_1) s_{12}c_{12}s_{13}\;,\nonumber\\
&&M_{23}= M_{32}= -{1\over 2} \left ( m_1(s^2_{12} + c^2_{12}s^2_{13}) + m_2(c^2_{12} + s^2_{12}s_{13}^2) - m_3 c^2_{13})\right )\;,\\
&&M_{12, 13}=M_{21, 31} = \pm {1\over \sqrt{2}}(m_2-m_1) s_{12}c_{12}c_{13} + i {s_{13}c_{13}\over \sqrt{2}} \left (m_1c_{12}^2 + m_2s^2_{12} + m_3\right )\;.\nonumber
\end{eqnarray}
Note that in the above the masses $m_i$ in general have Majorana phases. Multiplying a diagonal matrix $dig(1, 1, -1)$ from left and right to $M_\nu$ in Eq.\ref{mass}, one can obtain a new neutrino mass matrix $\tilde M_\nu = (\tilde M_{ij})$ in the form which has been discussed in the literature with~\cite{he1}
\begin{eqnarray}\label{mass-1}
&&\tilde M_{11} = M_{11}\;,\;\; \tilde M_{22,33} = M_{22,33}\;,\;\; \tilde M_{23} = - M_{23}\;,\\
&&\tilde M_{12, 13} = {1\over \sqrt{2}}(m_2-m_1) s_{12}c_{12}c_{13} \pm i {s_{13}c_{13}\over \sqrt{2}} \left (m_1c^2_{12}+ m_2s^2_{12} + m_3 \right )\;.\nonumber 
\end{eqnarray}

The resulting mass matrix indeed has a simple form with a high symmetric structure. It has been shown that the form of the above mass matrix can be obtained by imposing a symmetry transformation, for example for $m_i$ being real, with $e\to e$ and $\mu - \tau$ interchange with CP conjugation, the Grimus-Lavoura (GL) symmetry~\cite{GL-symmetry}, to obtain the mass matrix in Eq.\ref{mass-1}.
The $A_4$ symmetry for neutrino model is one of the interesting way to realize the matrix even when the neutrino masses have Majorana phases~\cite{he1,he2,ma1}. 

The statement for neutrino mixing pattern, such that the mixing angle $\theta_{23}$ 
rotating the second and third generation to be $\pi/4$ and the CP violating phase to be $-\pi/2$, is, 
of course, mixing parameterization convention dependent. 
When using a different convention, the situation may change. In the quark sector, the change of the values for mixing angles and CP violating phase has been pointed out some time ago~\cite{gerard}.
Therefore one may ask if the PDG parameterization is the unique one, with $\theta_{23} = \pi/2$ and $\delta_{PDG} = -\pi/2$, to describe the mixing matrix, even though the physical effects should not change~\cite{werner}. In the following we analyze the situation in details.
\\

\noindent
{\bf 2. Convention Dependence of Mixing Angles and Phase}

To answer the question whether mixing angle $\theta_{23}$ 
rotating the second and third generation to be $\pi/4$ and the CP violating phase to be $-\pi/2$ is unique or not, let us make a specific calculation using another well known parameterization, the original Kobayashi-Maskawa (KM) parameterization for quark mixing, given by~\cite{KM}
\begin{eqnarray}
V_{KM} = \left (
\begin{array}{ccc}
c_1&\;\;-s_1c_3&\;\;-s_1s_3\\
s_1c_2&\;\;c_1c_2c_3-s_2s_3 e^{i\delta_{KM}}&\;\;c_1c_2s_3+s_2c_3 e^{i\delta_{KM}}\\
s_1s_2&\;\;c_1s_2c_3+c_2 s_3 e^{i\delta_{KM}}&\;\;c_1s_2s_3-c_2c_3 e^{i\delta_{KM}}
\end{array}
\right )\;,
\end{eqnarray}
where $s_i = \sin\theta_i$ and $c_i = \cos\theta_{i}$.

The mixing angles and CP violating phase in KM parameterization can be expressed as functions of the parameters in the PDG parameterization by requiring the absolute values of the elements $V_{ij}$ in both of them to be the same and also
the Jarlskog parameter~\cite{jarlskog} $J$ to be the same since they are convention independent quantities. We have
\begin{eqnarray}
&&\vert V_{e1}\vert = c_1 = c_{12}c_{13}\;,\;\; \vert V_{e3} \vert = s_1s_3 = s_{13}\;, \nonumber\\
&&\vert V_{\mu 1}\vert^2 = (s_1c_2)^2 = (s_{12}c_{23})^2 + (c_{12}s_{23}s_{13})^2 + 2 s_{12}c_{12}s_{23}c_{23}s_{13} \cos\delta_{PDG}\;,\\
&&J = c_1s_1^2 s_2c_2 s_3 c_3\sin\delta_{KM} = s_{12}c_{12}s_{23}c_{23}s_{13}c_{13}^2\sin\delta_{PDG}\;.\nonumber
\end{eqnarray}

Setting $s_{23} = c_{23} = 1/\sqrt{2}$ and $\delta_{PDG} = -\pi/2$, we have
\begin{eqnarray}
c_{2} = s_{2} = {1\over \sqrt{2}}\;,\;\; \delta_{KM} = -{\pi \over 2}\;,\;\;s_{1} = \sqrt{1-c^2_{12} c^2_{13}}\;,\;\;s_3 = {s_{13}\over \sqrt{1-c^2_{12}c^2_{13}}}\;.
\end{eqnarray}
It is interesting to note that in the original KM parameterization, one of the mixing angles is also $\pi/4$ and the CP violating phase is also $-\pi/2$. 
It is then clear that the PDG parameterization is not unique in this view point. It should be kept in mind  that such degeneracy exists when making statements about neutrino mixing pattern.

One might then ask whether it is true that any parameterization also has one mixing angle to be $\pi/4$ and the CP violating phase $\delta$ is such a way that
$\sin \delta$ to be maximal. If not, how many different types of parameterizations have the above properties? 

We now use the classification of different types of parameterizations for mixing matrix in Ref. \cite{xing} to analyze the situation. It has been argued that when parameterizing the mixing matrix in terms of rotations around 1-2 plane by $R_{12}(\alpha)$, 1-3 plane by $R_{13}(\beta)$ and 2-3 plane by $R_{23}(\gamma)$ in flavor spaces depending on the ordering and also whether to repeat one of the rotations, there are 9 independent fashions. They are listed in Table 1 of Ref.~\cite{xing} .
Here $R_{ij}(\alpha)$ are given by
\begin{eqnarray}
R_{12}(\alpha) = \left (
\begin{array}{ccc}
c_\alpha&\;\;s_\alpha&\;\;0\\
-s_\alpha&\;\;c_\alpha&\;\;0\\
0&\;\;0&\;\;1
\end{array}
\right )\;,\;\;
R_{31}(\alpha) = \left (
\begin{array}{ccc}
c_\alpha&\;\;0&\;\;s_\alpha\\
0&\;\;1&\;\;0\\
-s_\alpha&\;\;0&\;\;c_\alpha
\end{array}
\right )\;,\;\;
R_{23}(\alpha) = \left (
\begin{array}{ccc}
1&\;\;0&\;\;0\\
0&\;\;c_\alpha&\;\;s_\alpha\\
0&\;\;-s_\alpha&\;\;c_\alpha
\end{array}
\right )\;.
\end{eqnarray}
When CP violating phase is introduced, one replaces the above 
rotation matrices by $R_{ij}(\alpha, \phi)$ which is obtained by replacing the entries ``1'' in $R_{ij}(\alpha)$ by $e^{-i\phi}$. Here $\phi$ is the CP violating phase.

The 9 independent parameterizations are given by
\begin{eqnarray}
&&V_{P_1} =R_{12}(\theta)R_{23}(\sigma, \phi) R^{-1}_{12}(\theta^\prime)\;,\;\;
V_{P_2} =R_{23}(\sigma)R_{12}(\theta, \phi) R^{-1}_{23}(\sigma^\prime)\;,\nonumber\\
&&V_{P_3} =R_{23}(\sigma)R_{31}(\tau, \phi) R_{12}(\theta)\;,\;\;\;\;V_{P_4} = R_{12}(\theta) R_{31}(\tau, \phi)R^{-1}_{23}(\sigma)\;,
\nonumber\\
&&V_{P_5} = R_{31}(\tau)R_{12}(\theta, \phi)R^{-1}_{31}(\tau^\prime)\;,\;\;V_{P_6} = R_{12}(\theta) R_{23}(\sigma, \phi) R_{31}(\tau)\;,\\
&&V_{P_7} = R_{23}(\sigma) R_{12}(\theta, \phi) R^{-1}_{31}(\tau)\;,\;\;\;V_{P_8} = R_{31}(\tau) R_{12}(\theta,\phi) R_{23}(\sigma)\;,\nonumber\\
&&V_{P_9} = R_{31}(\tau) R_{23}(\sigma,\phi) R^{-1}_{12}(\theta)\;. \nonumber \label{d-parametrization}
\end{eqnarray}

Each of them has some advantages in practical use and has been studied in the literature~\cite{xing}. In the following we will use these forms of parameterization to discuss which ones can have one of the mixing angles to be $\pi/4$ and the CP violating phase to be $-\pi/2$ to produce the mass matrix $M_\nu$ in Eq.\ref{mass}.

In the list, with redefinition of appropriate phases, $V_{P_2}$ and $V_{P_3}$  are equivalent to $V_{KM}$ and $V_{PDG}$, respectively. 
$V_{KM}$ from $V_{PDG}$ can be obtained from $V_{P_3}$ and $V_{P_2}$ following the rules below,  
\begin{eqnarray}
&&\mbox{$V_{KM}$ from $V_{P_2}$}:\;\;\;\mbox{identify}\;\;\delta_{KM} = \pi - \phi\;,\;\; \theta_1 = - \theta\;, \;\;\theta_2= - \sigma\;,\;\;\theta_3 = - \sigma^\prime\;. \nonumber\\
&&\mbox{$V_{PDG}$ from $V_{P_3}$}:\;\;V_{PDG} = diag(1, e^{i\phi}, e^{i\phi})\cdot V_{P_3} \cdot diag(1,1,e^{-i\phi})\;,\\
&&\hspace{2.9cm}\mbox{identify}\;\; \theta_{12} = \theta\;,\;\;\theta_{23} = \sigma\;,\;\;\theta_{13} = \tau\;, \;\; \delta_{PDG} = \phi\;.\nonumber
\end{eqnarray}
We will keep the notations used before for $V_{KM}$ and $V_{PDG}$ in our later discussions. 

There are parameterizations in which the Majorana phases are explicitly kept in the mixing matrix, such as the the symmetrical parameterization advocated in Ref.\cite{symmetric} . In such parameterizations are particularly informative for physical processes which are sensitive to Majorana phases, such as neutrinoless double beta decays, so that once measured, one knows certain combinations of the Majorana phases. Such parameterizations are also convenient in discussing phase invariance in neutrino sector when Majorana phases are considered \cite{majorana-phase}. In  our discussions, only the Dirac CP violating phase $\delta$ and the mixing angles from neutrino oscillations are involve, which do not carry information about Majorana phases, the specific parameterization for Majorana phases will not affect our results. The symmetrical parameterization is actually equivalent to $V_{P_3}$ as far as properties related to the mixing angles and Dirac CP violating phase are concerned. The same is true for all other parameterizations listed in eq.\ref{d-parametrization} in relation to their extended parameterizations where Majorana phases appear in certain form. Without loss of generality, we will just need to analyze the parameterization in eq.\ref{d-parametrization}.

We have seen that both the above parameterizations have one of the mixing angles, the rotating angle in 2-3 plane, to be $\pi/4$ and the CP violating phase to be $-\pi/2$. Among the 9 independent parameterizations, we find $V_{P_7}$ also has the above property, when $\theta_{23} = \pi/4$ and $\delta = -\pi/2$, we have
\begin{eqnarray}
s_\sigma = c_\sigma = {1\over \sqrt{2}}\;,\;\;\phi = -{\pi\over 2}\;,\;\; s_\theta = s_{12}c_{13}\;,\;\;s_\tau = {s_{13} \over (1- s_{12}^2 c_{13}^2 )^{1/2}}\;.
\end{eqnarray}

Specific calculations for other cases, we find that $V_{P_{1, 4, 5, 6, 8, 9}}$ do not have the properties that one of the mixing angles is $\pi/4$ and the CP violating phase $\phi$ to be $- \pi/2$ (or $\pi/2$. The sign of $\phi$ can be changed by phase redefinition).  For $\theta_{23}=\pi/4$ and $\delta_{PDG}$ in $V_{PDG}$, we have
\begin{eqnarray}
&&V_{P_1}:\;\;s_{\sigma} = {1\over \sqrt{2}} \sqrt{1+s^2_{13}}\;,\;\;s_\theta =  {\sqrt{2} s_{13}\over \sqrt{1+s^2_{13}}}\;,\;\; s_{\theta^\prime} = {\sqrt{1-c^2_{12}c^2_{13} }\over \sqrt{1+s^2_{13}} }\;,\nonumber\\
&&\hspace{1.1cm}\sin\phi = -{s_{12}c_{12}  (1+s^2_{13}) \over \sqrt{(1-c^2_{12}c^2_{13})(1-s^2_{12}c^2_{13})}}\;.\nonumber\\
&&V_{P_4}:\;\;s_\tau = {1\over \sqrt{2}}\sqrt{1-c_{12}^2c^2_{13}}\;,\;\;s_\theta = {\sqrt{1-c^2_{12}c^2_{13}}\over \sqrt{1+c^2_{12}c^2_{13}}}\;,\;\;s_\sigma = {\sqrt{1-s^2_{12}c^2_{13}}\over \sqrt{1+c^2_{12}c^2_{13}}}\;,\nonumber\\
&&\hspace{1.1cm}\sin\phi = - {s_{12}s_{13}(1+c_{12}^2c^2_{13})\over (1-c^2_{12}c^2_{13})\sqrt{1-s^2_{12}c^2_{13}}}\;.\nonumber\\
&&V_{P_5}:\;\;s_\tau = {\sqrt{1-s_{12}^2c^2_{13}}\over \sqrt{1+s^2_{12}c^2_{13}}}\;,\;\;s_\theta = {1\over \sqrt{2}}{\sqrt{1+s^2_{12}c^2_{13}}}\;,\;\;s_{\tau^\prime} = {c_{13}\over \sqrt{1+s^2_{12}c^2_{13}}}\;,\\
&&\hspace{1.1cm}\sin\phi = - {c_{12}s_{13}(1+s_{12}^2c^2_{13})\over \sqrt{1-c^2_{12}c^2_{13}}(1-s^2_{12}c^2_{13})}\;.\nonumber\\
&&V_{P_6}:\;\;s_\tau = { \sqrt{1-c_{12}^2c^2_{13}}\over \sqrt{1+s^2_{12}c^2_{13}}}\;,\;\;s_\theta = {\sqrt{2}s_{12}c_{13} \over \sqrt{1+s^2_{12}c^2_{13}}}\;,\;\;s_{\sigma} = {1\over \sqrt{2}} \sqrt{1-s^2_{12}c^2_{13}}\;,\nonumber\\
&&\hspace{1.1cm}\sin\phi = - {c_{12}s_{13}(1+s_{12}^2c^2_{13})\over \sqrt{ 1-c^2_{12}c^2_{13}}(1-s^2_{12}c^2_{13})}\;.\nonumber\\
&&V_{P_8}:\;\;s_\tau = { \sqrt{1-c_{12}^2c^2_{13}}\over \sqrt{1+c^2_{12}c^2_{13}}}\;,\;\;s_\theta = {1\over \sqrt{2}} \sqrt{1-c^2_{12}c^2_{13}}}\;,\;\;s_{\sigma} = {c_{13}\over  \sqrt{1+c^2_{12}c^2_{13}}\;,\nonumber\\
&&\hspace{1.1cm}\sin\phi = - {s_{12}s_{13}(1+c_{12}^2c^2_{13})\over (1-c^2_{12}c^2_{13})\sqrt{1-s^2_{12}c^2_{13}}}\;.\nonumber\\
&&V_{P_9}:\;\;s_\tau = {\sqrt{2}s_{13}\over \sqrt{1+s^2_{13}}}\;,\;\;s_\theta = {\sqrt{1-c^2_{12}c^2_{13}} \over \sqrt{1+s^2_{13}}}\;,\;\;s_{\sigma} = {1\over \sqrt{2}} c_{13}\;,\nonumber\\
&&\hspace{1.1cm}\sin\phi = - {s_{12}c_{12}(1+s^2_{13}) \over \sqrt{1-c^2_{12}c^2_{13}}\sqrt{1-s^2_{12}c^2_{13}}}\;.\nonumber
\end{eqnarray}

Note that none of the mixing angles in $V_{P_{1,4,5,6,8,9}}$ is dictated to be $\pi/4$ and $\phi$ is not automatically $\pm\pi/2$ neither. One should be clear that this is not to say that these parameterizations cannot fit data if it turns out that experiments will determine $\theta_{23} = \pi/4$ and $\delta_{PDG} = -\pi/2$, but just means that the mixing angles and the CP violating phase are not necessarily $\pi/4$ and $-\pi/2$, respectively. In the following we will take neutrino mixing parameters with 
\begin{eqnarray}
s_{23} = 1/\sqrt{2}\;,\;\;\delta_{PDG} = -\pi/2\;,\;\;s_{12}=0.545\;,\;\;s_{13} = 0.147\;, \label{test}
\end{eqnarray}
allowed by data for both NH and IH cases, to illustrate the point by obtaining the parameters in $V_{P_1}$. We have 
\begin{eqnarray}
&&s_\sigma = 0.715\;,\;\; s_\theta = 0.206\;,\;\;s_{\theta^\prime} = 0.553\;,\;\;\sin\phi = -0.992\;.
\end{eqnarray}
The angles and phase are different than those in the PDG parameterization, but all are in physical regions which produces the same physical effect as that provided by the PDG parametrization.\\

\noindent
{\bf 3. Discussions and Conclusions}

From the list of the independent parameterizations, we notice that for those parameterizations, $V_{P_{2,3,7}}$, the rotations start with $R_{23}(\alpha)$ from the left of the product for $V_{PMNS}$ will result in one of the mixing angle is dictated to be $\pi/4$ and the CP violating phase to be $-\pi/2$. What are then the differences between the 3 parameterizations? They show up in the numerical values of the angles in rotations in the other two planes.  Taking the values in Eq.\ref{test} for the angles and phase in PDG parameterization, we have
\begin{eqnarray}
&&V_{PDG}:\;\;\;s_{12} = 0.545\;,\;s_{13} = 0.147\;,\nonumber\\
&&V_{KM}:\;\;\;\;s_1 = 0.559\;,\;\; s_3 = 0.263\;,\\
&&V_{P_7}: \;\;\;\;\;\;s_\theta = 0.539\;,\;\; s_{\tau} = 0.175\;.\nonumber
\end{eqnarray}

For the other parameterizations listed in Eq.\ref{d-parametrization}, we have their angles and phase given by
\begin{eqnarray}
V_{P_1}:&&\;s_\sigma = 0.715\;,\;\; s_\theta = 0.206\;,\;\;s_{\theta^\prime} = 0.553\;,\;\;\sin\phi = -0.992\;,\nonumber\\
V_{P_4}:&&\;s_\tau = 0.395\;,\;\; s_\theta = 0.430\;,\;\;s_\sigma = 0.648\;,\;\;\sin\phi = -0.514\;,\nonumber\\
V_{P_5}:&&\;s_\tau = 0.741\;,\;\; s_\theta = 0.803\;,\;\;s_{\tau^\prime} = 0.871\;,\;\;\sin\phi = -0.401\;,\nonumber\\
V_{P_6}:&&\;s_\tau = 0.492\;,\;\; s_\theta = 0.671\;,\;\;s_\sigma = 0.596\;,\;\;\sin\phi = -0.401\;,\\
V_{P_8}:&&\;s_\tau = 0.430\;,\;\; s_\theta = 0.395\;,\;\;s_\sigma = 0.761\;,\;\;\sin\phi = -0.514\;,\nonumber\\
V_{P_9}:&&\;s_\tau= 0.206\;,\;\; s_\theta = 0.553\;,\;\;s_\sigma = 0.699\;,\;\;\sin\phi = - 0.992\;.\nonumber
\end{eqnarray}

One clearly sees that the values for the mixing angles and the CP violating phase can be very different for conventions adopted for
the parameterizations. When coming to physical observables, they all have the same effects. But sometimes, a particular convention
of the parameterization is convenient and helps to reveal certain structures, such the G-L symmetry in neutrino mass matrix is
more transparent to be understood in the PDG parameterization. Nevertheless, one should not over state the usefulness of a certain parameterization. One should give the specific parameterization convention when making statements about values for mixing angles and phase.

\section*{Acknowledgments}
This work was supported in part by Key Laboratory for Particle Physics, Astrophysics and Cosmology, Ministry of Education, and Shanghai Key Laboratory for Particle Physics and Cosmology (Grant No. 15DZ2272100), and in part by the NSFC (Grant Nos. 11575111 and 11735010). XGH was supported in part by the MOST (Grant No. MOST 106-2112-M-002-003-MY3 ).

\end{document}